\documentclass[aps,prl,twocolumn,showpacs]{revtex4}
\usepackage[T1]{fontenc}
\usepackage[latin1]{inputenc}
\usepackage{graphics}
\usepackage{amsmath}

\newcommand{\beq}[1][]{\begin{equation}\label{#1}}
\newcommand{\eeq}{\end{equation}}

\newcommand{\bxi}{\ensuremath{\boldsymbol{\xi}}}
\newcommand{\A}{\ensuremath{\boldsymbol{A}}}
\newcommand{\B}{\ensuremath{\boldsymbol{B}}}

\makeatletter

\providecommand{\LyX}{L\kern-.1667em\lower.25em\hbox{Y}\kern-.125emX\@}

\makeatother

\begin{document}

\title{$q$-Breathers and the Fermi-Pasta-Ulam Problem}
\author{S.~Flach$^1$, M.~V.~Ivanchenko$^2$, and O.~I.~Kanakov$^2$}

\affiliation{$^1$ Max-Planck-Institut f\"ur Physik komplexer
Systeme, N\"othnitzer Str. 38, D-01187 Dresden, Germany
\\
$^2$ Department of Radiophysics, Nizhny Novgorod University,
Gagarin Avenue, 23, 603950 Nizhny Novgorod, Russia }

\begin{abstract}
The Fermi-Pasta-Ulam (FPU) paradox consists of
the nonequipartition of energy among normal modes of a weakly
anharmonic atomic chain model. In the harmonic limit each normal
mode corresponds to a periodic orbit in phase space and is
characterized by its wave number $q$. We continue normal modes
from the harmonic limit into the FPU parameter regime and obtain
persistence of these periodic orbits, termed here $q$-Breathers (QB).
They are characterized by time periodicity, exponential
localization in the $q$-space of normal modes and linear
stability up to a size-dependent threshold amplitude. 
Trajectories computed in the original FPU setting are
perturbations around these exact QB solutions. The QB concept
is applicable to other nonlinear lattices as well.

\end{abstract}


\maketitle

Fifty years ago E.~Fermi, J.~Pasta, and S.~Ulam (FPU) published
their celebrated report on thermalization of arrays of
particles connected by weakly nonlinear strings \cite{fpu},
bringing forth a fundamental physical and mathematical problem
of the energy equipartition and ergodicity in nonlinear systems.
Series of numerical simulations showed that energy, initially
placed in a low frequency normal mode of the linear problem with
a frequency $\omega_q$ and a corresponding wave number $q$,
stayed almost completely locked within a few neighbour modes,
instead of being distributed among all modes of the system.
Moreover, recurrence of energy to the originally excited mode
was observed.

Much effort has since been expended 
to understand and explain the FPU results
(see \cite{Ford,chaosfpu,Izrailev} for
reviews). Two major approaches were developed. The first one,
taken by N.~Zabusky and M.~Kruskal, was to analyze dynamics of the
nonlinear string in the continuum limit, which led to a
pioneering observation of solitary waves \cite{Zabusky}. 
The second approach, followed by F.~Izrailev and B.~Chirikov pointed to
the existence of a "stochasticity threshold" in the original FPU
system \cite{Izrailev_Chirikov}. For strong nonlinearity (or
simply large energies) the overlap of nonlinear resonances
\cite{Chirikov} leads to strong dynamical chaos, destroying the
FPU recurrence and insuring fast convergence to thermal
equilibrium. If nonlinearity is below a (size-dependent)
"stochasticity threshold", the dynamics of the chain remains
similar to that of the unperturbed (linear) system for large time
scales. Later studies \cite{deLuca,deLuca2002} showed that the
"local" dynamics of four consecutive low-frequency modes may
become substantially chaotic, while almost all initial energy
stays localized in these modes during the time of
computation. The redistributed mode energies fall
exponentially with increasing mode numbers in this regime
(coined "weak chaos") and the energy flow to higher
frequency modes was argued to be exponentially slow (due to
Arnold diffusion).

The results obtained through this second approach lead one to
formulate some important questions. Firstly, since the dynamics
below the stochasticity threshold is localized in $q$-space for
long times, do {\it time-periodic} trajectories with almost all
energy locked in a single mode for {\it infinite times}, coined
{\it $q$-Breathers} (QB), exist, and are they close to the ones
studied by FPU? Secondly, are the stability thresholds of
such QBs related to the various stochasticity thresholds
mentioned above? And finally, is the concept of QBs applicable
also for other spatially extended nonlinear lattices,
including generalizations to higher lattice dimensions?
A strong motivation for this study is the fact
that in the $q$-space representation we deal with oscillators
which are uncoupled in the limit of small amplitudes, and,
moreover, with frequencies being different for each mode.
Nonlinearity induces coupling between oscillators. That is
reminiscent of the case of {\it discrete breathers} (DB), 
that are time-periodic and spatially localized excitations on
networks of interacting identical anharmonic oscillators, which
survive continuation from the trivial limit of zero coupling
\cite{mackayaubry}. Notably, DBs exist also in FPU lattices
\cite{DBs}.

The FPU system is a chain of $N$ equal masses coupled by nonlinear
strings with the equations of motion containing quadratic (the
$\alpha$-model)
\begin{equation}
\label{eq1}
 \ddot{x}_n=(x_{n+1}-2x_n+x_{n-1})\\
+\alpha[(x_{n+1}-x_n)^2-(x_n-x_{n-1})^2]
\end{equation}
or cubic (the $\beta$-model)
\begin{equation}
\label{eq2}
 \ddot{x}_n=(x_{n+1}-2x_n+x_{n-1})
+\beta[(x_{n+1}-x_n)^3-(x_n-x_{n-1})^3]
\end{equation}
interaction terms, where $x_n$ is the displacement of the $n$-th
particle from its original position, and fixed boundary
conditions are taken $x_0=x_{N+1}=0$. A canonical transformation
$x_n(t)=\sqrt{\frac{2}{N+1}}\sum\limits_{q=1}^N
Q_q(t)\sin{\left(\frac{\pi q n}{N+1}\right)}$ takes into the
reciprocal wavenumber space with $N$ normal mode coordinates
$Q_q(t)$. The equations of motion then read
\begin{equation}
\label{eq6}
 \Ddot{Q}_q+\omega_q^2 Q_q=-\frac{\alpha}{\sqrt{2(N+1)}}\sum\limits_{i,j=1}^N
 A_{q,i,j}Q_iQ_j
\end{equation}
for the FPU-$\alpha$ chain (\ref{eq1}) and
\begin{equation}
\label{eq7}
 \Ddot{Q}_q+\omega_q^2 Q_q=-\frac{\beta}{2(N+1)}\sum\limits_{i,j,m=1}^N
 C_{q,i,j,m}Q_iQ_jQ_m
\end{equation}
for the FPU-$\beta$ chain (\ref{eq2}), where
$\omega_q=2\sin{(\pi q/2(N+1))}$ are the normal mode frequencies, and
$A_{q,i,j}$ and $C_{q,i,j,m}$
are coupling coefficients \cite{deLuca}.
For small amplitude excitations the nonlinear terms
in the equations of motion can be neglected, and according
to (\ref{eq6}) and (\ref{eq7}) the $q$-oscillators get decoupled,
each conserving the energy
$E_q=\frac{1}{2}\left(\dot{Q}_q^2+\omega_q^2 Q_q^2\right)$
in time. Especially, we may consider the excitation of only one
of the $q$-oscillators, i.e. $E_{q} \neq 0$ for $q\equiv q_0$ only.
Such excitations are trivial
time-periodic and $q$-localized solutions (QBs)
for $\alpha=\beta=0$.

Let us consider the $\beta$-model and
choose first $\beta=0$, excite a normal mode with $q=q_0$ to the
energy $E_{q_0}=E$ and let all other $q$-oscillators be at
rest. With that we arrive at a unique periodic orbit in the
phase space of the FPU model. In order to guarantee continuation of this
periodic orbit into the case $\beta \neq 0$ and following
\cite{mackayaubry} we conclude, that it is enough to request 
the nonresonance condition $n \omega_{q_0} \neq \omega_{q \neq q_0}$
which is the generic case for a finite system size $N$
(here $n$ is an integer) (see also \cite{oldstuff}.
We expect that the
orbit will stay localized in $q$-space at least up to
some critical nonzero value of  $\beta$.
The same argumentation can be applied to the
$\alpha$-model \cite{symmanifolds}.

We compute QBs as well as their
Floquet spectrum numerically  using well developed computational tools \cite{DBs},
and compare the results with analytical predictions, derived by means of asymptotic expansions.
As a zero-order approximation for the numerical computation we take the
$q_0$-th linear mode:
%\begin{equation}
%\label{eq8}
$ x_n(t)=\sqrt{\frac{2}{N+1}}Q_{q_0}(t)\sin{\left(\frac{\pi q_0
n}{N+1}\right)}$.
%\end{equation}
For the $\beta$-model 
the initial conditions are $Q_q(t=0)=0$, and $\dot{Q}_{q_0}(t=0)=
\sqrt{2E-\sum_{q \neq q_0}\dot{Q}_q^2(t=0)}$. We map the space
of $\vec{y}\equiv \{ \dot{Q}_q\}$ 
onto itself by integrating the initial condition
up to the time when $Q_{q_0}(t)=0$ again: 
$\vec{y}^{n+1}=\vec{\mathcal{F}}(\vec{y}^n)$. 
A periodic orbit is a fixed point of that map.
The vector function $\vec{\mathcal{G}}=\vec{\mathcal{F}}(\vec{y})
-\vec{y}$ is used to calculate the Newton matrix
$\mathcal{N}=\partial{\mathcal{G}(\vec{y})_i}/\partial y_j$.
The iteration procedure $\vec{y}\ '
=\vec{y}-\mathcal{N}^{-1}\vec{\mathcal{G}}(\vec{y})$ continues
until the required accuracy $\varepsilon$ is obtained:
$||\vec{\mathcal{F}}(\vec{y})-\vec{y}||/||\vec{y}||<\varepsilon$
(we have varied $\varepsilon$ from $10^{-5}$ to $10^{-8}$),
where $||\vec{y}||=\max\{|y_i|\}$.
For the $\alpha$-model we used a modified scheme
choosing $x_s(t=0)=0$ where $s=[(N+1)/2q_0]$ corresponds
to the antinode of the mode $Q_{q_0}$. We map the phase space
$\vec{r}$ (excluding $x_s$) onto itself integrating until $x_s(t)=0$ again.
With the above notations we use a Gauss method to solve the
equations $\vec{\mathcal{G}}(\vec{r})={\mathcal{N}}(\vec{r} - \vec{r}\ ')$
for the new iteration $\vec{r}\ '$ and do final corrections to
adjust the correct energy $E$.
To compute the linear stability of the found QB,
we linearize the phase space flow around it, and map that flow
onto itself by integrating over one period of the QB.
The corresponding symplectic Floquet matrix 
can be computed numerically and subsequently diagonalized.
If all eigenvalues $\mu$ have absolute value one, the QB is stable,
otherwise it is unstable \cite{DBs}.

\begin{figure}[t]
{\centering
\resizebox*{0.90\columnwidth}{!}{\includegraphics{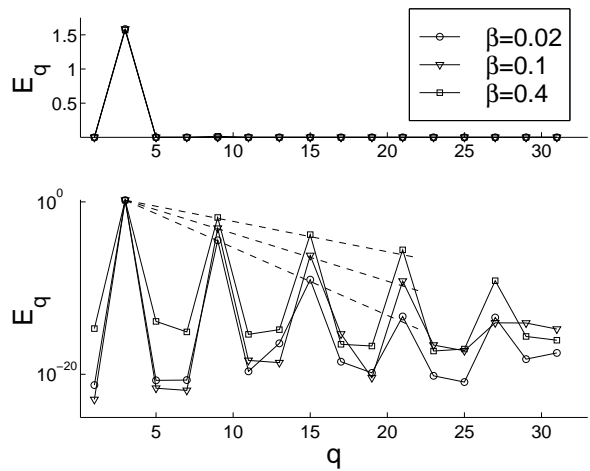}}}
{\caption{Energy distributions between $q$-modes in QBs for
different nonlinear coupling coefficients $\beta$ versus $q$ in linear and log
scales with analytical estimations of the
QBs exponential localization (dashed lines). Parameters are
$E=1.58$, $q_0=3$, $N=32$. Only odd modes are shown (see text).
The symbols for $q \neq 3,9,15,21,27$ represent upper bounds,
the real mode energies might be even less.
Note that QBs persist even far beyond the
stability threshold (see Fig.\ref{fig2}).
}\label{fig1}}
\end{figure}

\begin{figure}[t]
{\centering
\resizebox*{0.90\columnwidth}{!}{\includegraphics{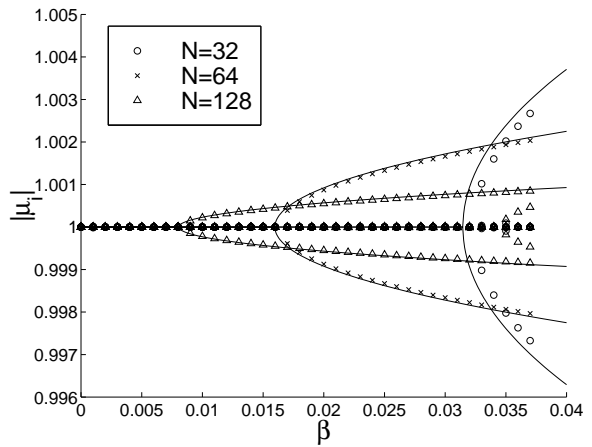}}}
{\caption{Absolute values of Floquet multipliers $|\mu_i|$ of QBs
with the energy $E=1.58$ and $q_0=3$ and different $N$ versus
$\beta$. Symbols: numerical results, lines: analytical results.}
\label{fig2}}
\end{figure}

First, we apply our method to the $\beta$-model with $q_0=3$ and
$E=1.58$ which is very close to the value 1.5 chosen in
\cite{deLuca}. We obtain QBs which are exponentially localized in
$q$-space (Fig.\ref{fig1}). The smaller $\beta$, the faster is the
decay of the energy distribution with increasing wave number $q$.
Note, that due to the parity symmetry of the $\beta$-model
(Eq.(\ref{eq7}) is invariant under $x_n \rightarrow -x_n$ for all
$n$) only odd $q$-modes are excited by the $q_0=3$ mode and get
coupled \cite{bushes}. In Fig.\ref{fig2} we plot the absolute
values of the Floquet eigenvalues of the computed QBs versus
$\beta$ for different system sizes $N$. QBs are stable for
sufficiently weak nonlinearities (all eigenvalues have absolute
value 1). When $\beta$ exceeds a certain threshold two eigenvalues
get absolute values larger than unity (and, correspondingly,
another two get absolute values less than unity) and a QB becomes
unstable. Remarkably, unstable QBs can be traced far beyond the
stability threshold, and moreover, they retain their exponential
localization in $q$-space (Fig.\ref{fig1}). As $\beta$ is
increased further, new bifurcations of the same type are observed.

\begin{figure}[t]
{\centering
\resizebox*{0.90\columnwidth}{!}{\includegraphics{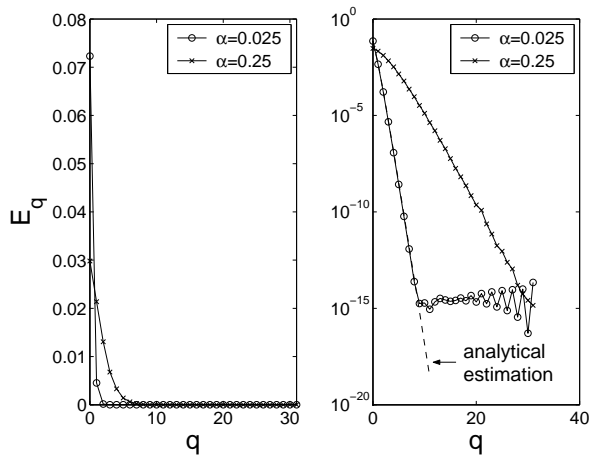}}}
{\caption{Stable QB solutions for $\alpha=0.025$ and $\alpha=0.25,
E=0.077, N=32$, and $q_0=1$; the latter corresponds to the
original numerical FPU-$\alpha$ study \cite{fpu}. An analytical
estimation of the QB exponential localization in case
$\alpha=0.025$ is shown with a dashed line.}\label{fig3}}
\end{figure}

By an asymptotic expansion of the solution to \eqref{eq7} in
powers of the small parameter $\sigma=\beta/(N+1)$ we estimate
the shape of a QB solution $\hat{Q}_i(t)$ localized in the
mode $q_0$. The energies of the modes $q_0$,
$3q_0$,\dots,$(2n+1)q_0$,\dots\ are given by
\begin{equation}
\label{Eq21}
E_{(2n+1)q_0}=\lambda^{2n}E_{q_0}\;,\;
\lambda=\frac{3\beta E_{q_0}(N+1)}{8\pi^2 q_0^2}\;,\;
\end{equation}
%\end{subequations}
up to a an error of the order $(2n+1)^2 q_0^2/N^2$.
Dashed lines in Fig.\ref{fig1} are obtained using (\ref{Eq21})
and show very good agreement with the numerical results.

Using standard secular perturbation techniques we approximate
the frequency $\Omega$ of the QB solution  as $ \Omega =
\omega_{q_0}(1+\frac{9\beta E_{q_0}}{8(N+1)}
+O(\frac{\beta^2}{(N+1)^2}))\;. $ The instability threshold
observed in Fig.\ref{fig2} can be obtained analytically by
making a replacement $Q_i=\hat{Q}_i(t)+\xi_i$ in the equations
of motion \eqref{eq7} and linearizing the resulting equations
with respect to $\xi$:
\beq[Math]
\Ddot{\bxi}+\A\bxi+h(1+\cos2\Omega
t)\B\bxi+O(h^2)\bxi=0
\eeq
where $\bxi=(\xi_i)$ is a vector,
$\A=(\delta_{ij}\omega_i^2)$ is a diagonal matrix, $\B=(b_{ij})$
is a coupling matrix, and $h=3\beta E/2(N+1)$ is a small
parameter. We analyze parametric resonance in
\eqref{Math}, treating $h$ and $\Omega$ as independent
parameters. In the limit $h\to 0$ the equilibrium point $\bxi=0$
is stable for all values of $\Omega$ except for those which
satisfy $ \omega_k+\omega_l=2n\Omega \equiv 2n \Omega_{nkl} $
where $n\ge1$, and the modes $k$ and $l$ belong to the same
connected component of the coupling graph whose connectivity is
defined by the matrix \B.

We seek for a solution to \eqref{Math} at $\Omega=\Omega_{nkl}(1+\delta)$,
$\delta=O(h)$, in the form
$\bxi=\sum_{m=-\infty}^{+\infty} \boldsymbol{f}_m
e^{(i\Tilde{\omega}+z+2im\Omega)t}+c.c.
$
where
$\Tilde{\omega}=\omega_k(1+\delta)=-\omega_l(1+\delta)+2n\Omega$,
$\boldsymbol{f}_m$ are unknown complex vector amplitudes, and
$z=O(h)$ is a small unknown complex number.

The nearest primary resonance corresponds to $k=q_0-1$, $l=q_0+1$,
$n=1$ \cite{deLuca}. In the vicinity of the bifurcation point
the absolute values of the Floquet multipliers involved in
the resonance are obtained as
\beq[mults]
|\mu_{j_1 j_2}|=1\pm\frac{\pi^3}{4(N+1)^2}
\sqrt{R-1+O\left(\frac{1}{N^2}\right)}
\eeq
where $R=6\beta E(N+1)/\pi^2$. The bifurcation occurs at
$R=1+O(1/N^2)$. The result \eqref{mults} is plotted in
Fig.\ref{fig2} with solid lines for $N=32$, 64 and 128,
demonstrating good agreement with the numerical results. The
agreement improves with increasing $N$ \cite{corrections}.
The instability threshold for QB orbits (Fig.\ref{fig2}) which
is obtained analytically using the parameter $R$ (\ref{mults}),
coincides with the criterion of transition to weak chaos
reported by De Luca et al \cite{deLuca}.

We have used
one of the original parameter sets of the FPU$-\alpha$ study
$\alpha=0.25, E=0.077, N=32$ \cite{fpu} (and add to that
the case $\alpha=0.025$ as well for comparison) to find stable
exponentially localized QB modes with most of the energy
concentrated in $q_0=1$ (Fig.\ref{fig3}).
We use again an
asymptotic expansion of the solution to \eqref{eq6} in powers of
the small parameter $\rho=\alpha/\sqrt{2(N+1)}$ and obtain that the
energies of the
modes $q_0$, $2q_0$,\dots,$n q_0$, are given by
\begin{equation}
\label{Eq22} E_{n q_0}=\epsilon^{2n-2}n^2 E_{q_0}\;,\;
\epsilon=\frac{\alpha \sqrt{E_{q_0}^{(0)}}(N+1)^{3/2}}{\pi^2
q_0^2}\;,\;
\end{equation}
%\end{subequations}
The dashed line in Fig.\ref{fig3} is obtained using (\ref{Eq22})
in case $\alpha=0.025, E=0.077, N=32, q_0=1$ and shows very good
agreement with the numerical results \cite{higher}.

\begin{figure}[t]
{\centering
\resizebox*{0.90\columnwidth}{!}{\includegraphics{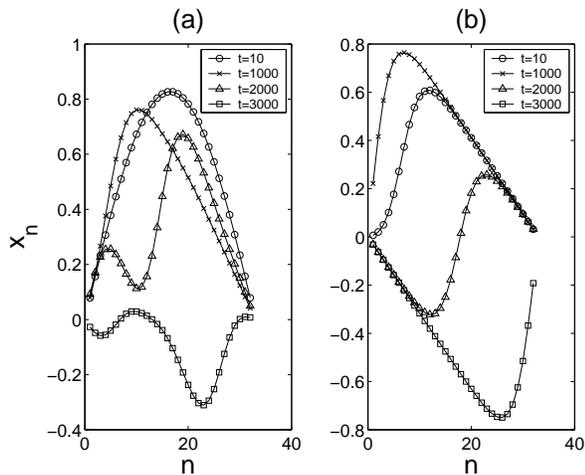}}}
{\caption{Snapshots of displacements (a) of the original FPU
trajectory for $\alpha=0.25, E=0.077, N=32$ \cite{fpu} and (b) of
the corresponding exact QB solution from Fig.\ref{fig3} taken at
different times. } \label{fig4}}
\end{figure}

How are QB-solutions related to the original FPU studies? A QB
requires specific initial conditions, which were not used in
earlier numerical studies.
However, nearby quasiperiodic solutions  are expected to
retain stability and
exponential localization of their QB generator for times long
compared to the QB period.
In Fig.\ref{fig4} we compare snapshots of displacements at
different times obtained for the original FPU trajectory in
\cite{fpu} and for the numerically exact QB solution from
Fig.\ref{fig3} for $\alpha=0.25$ and observe
similar evolution patterns. Moreover, we took a series of points
on a line which connected initial conditions of the FPU trajectory
($E_{q \neq 1} =0$) with the numerically exact QB solution from
Fig.\ref{fig3}. For each of these points we integrated the
corresponding trajectory and measured  the average deviation
$\Delta$ from the QB orbit. The dependence of $\Delta$ on the line
parameter turns out to be an almost linear one, starting from zero
when being very close to the QB orbit, and ending with a maximum
value when being close to the FPU trajectory. That supports the
expectation that the FPU trajectory is a perturbation of the QB
orbit. The FPU recurrence is gradually appearing with increasing
$\Delta$ and is thus
directly related to the regular motion of a slightly perturbed QB
periodic orbit, which we tested also numerically.

In conclusion, we report on the existence of $q$-breathers as exact
time-periodic low-frequency solutions in the nonlinear FPU system.
These solutions are exponentially localized in the $q$ space of
the normal modes and preserve stability for small enough
nonlinearity. They continue from their trivial counterparts for
zero nonlinearity at finite energy (in that limit they simply
correspond to one mode with wave number $q_0$ being excited, and
all the other modes being at rest). The stability threshold of QB
solutions coincides with the weak chaos threshold
in \cite{deLuca}.
Persistence of exact stable QB modes is
shown to be related to the FPU paradox. The FPU
trajectories computed 50 years ago are perturbations of the exact QB orbits.
Remarkably, localization in $q$-space persists even for parameters
when the QBs turn unstable. The concept of
stable QBs and their impact on the evolution of excitations in the
FPU system is expected to apply far beyond the stability threshold
of the QB solutions reported in the present work.
Generalizations to higher dimensional lattices and other
Hamiltonians are straightforward, due to the weak constraint imposed
by the nonresonance condition needed for continuation.
QBs can be also expected to contribute to peculiar dynamical features of
nonlinear lattices in thermal equilibrium, e.g.
the anomalous heat conductivity in
FPU lattices \cite{chaosfpu}.

We thank T. Bountis, F. Izrailev, V. Shalfeev 
and V. Zakrzewski
for stimulating discussions.
M.I. and O.K. appreciate the warm hospitality of the Max Planck Institute
for the Physics of Complex Systems. M.I. also acknowledges support of
the "Dynasty" foundation.


\begin{thebibliography}{99}
\bibitem{symmanifolds}
Note that we do not demand specific symmetries leading to
low-dimensional invariant manifolds in phase space, in contrast
to the discussion in \cite{bushes}.
Consequently, our QB solutions must not be embedded in such
subspaces.
\bibitem{corrections}
The subsequent bifurcations of the same type are associated with
next primary resonances of the type $k=q_0-m$, $l=q_0+m$, $n=1$,
$m=2,3,\dots$. The corresponding bifurcation points are located
at $R=m^2+O(1/N^2)$. Note, that higher-order resonances at
certain values of $m_0$ and $N$ may result in appearance of
narrow instability regions  within the region $0<R<1$.
\bibitem{higher} To describe the energy distribution for
$\alpha=0.25$,
a higher order perturbation theory needs to be developed.
\end{thebibliography}
\end{document}